\documentclass[letterpaper, 10 pt, conference]{ieeeconf}  

\IEEEoverridecommandlockouts                              

\usepackage{graphics} 
\usepackage{epsfig} 
\usepackage{mathptmx} 
\usepackage{times} 
\usepackage{amsmath} 
\usepackage{amssymb}  
\usepackage[mathcal]{eucal}
\usepackage{hyperref}

\DeclareMathOperator*{\minimize}{minimize}
\DeclareMathOperator*{\find}{find}
\DeclareMathOperator*{\subjectto}{subject \ to}

\title{\LARGE \bf
A Finite Element Method Approach for Trajectory Generation via Time-Optimal Control and Model Predictive Control Tracking}

\author{Jose A. Solano-Castellanos$^{1}$
\thanks{$^{1}$Jose A. Solano-Castellanos is with the Department of Mechanical Engineering at the Massachusetts Institute of Technology, 77 Massachusetts Ave, Cambridge, MA 02139, USA.
        {\tt\small jsolanoc@mit.edu}}}

\begin{document}

\maketitle
\thispagestyle{empty}
\pagestyle{empty}

\begin{abstract}

In this paper a framework for solving the time optimal control (TOC) using Galerkin's Weighted Residuals Method (GWRM) and Sequential Convex Programming (SCP) is proposed. The proposed method solves the two-point boundary value problem, avoiding the use of shooting methods that rely heavily on the appropriate initialization of the adjoint state and optimal time. Since TOC yields an open-loop controller, a Model Predictive Control (MPC) scheme is employed to track both the optimal trajectory and controller, allowing the system to reject disturbances. The approach is validated using the Dubins' car dynamics for optimal time trajectory generation.

\end{abstract}

\section{INTRODUCTION}

The time optimal control (TOC) problem consists of finding a control function $\mathbf{u}^{\star}(t)$ such that a system starting at the state $\mathbf{X}_0$ at time $t_0$ reaches a desired terminal set $\mathcal{X}_T$ in the least amount of time $T^{\star}$ while satisfying the dynamics of the system and possible constraints on the state and/or control inputs. Pontryagin’s Maximum Principle (PMP) provides a framework to solve the TOC problem which often is solved using shooting methods that are highly dependent on the initialization of the method. This can prevent the use of TOC in applications where one may want to execute a task in the least amount of time, but the waypoints are provided sequentially during operation, removing the possibility of computing the entire trajectory offline.

Finite Element Methods (FEM) present and alternative to shooting methods since they are able to deal with two-point boundary value problems by transforming the set of differential equations into a system of algebraic equations using a linear combination of basis functions over the domain and finding the modal coefficients of the solution that minimize the residual. Different from shooting methods, the result of the FEM are functions that approximate the solution over the domain, rather than discrete points at particular time steps.

Since the TOC provides an open-loop controller of the problem, to track the generated trajectory $\mathbf{X}^{\star}(t)$ for all $t\in[0, T^{\star}]$ and reject possible disturbances on the system, a Model Predictive Control (MPC) will be applied to produce a closed-loop controller around the nominal trajectory and controller. 

As a motivation for this approach, the dynamics of a Dubins’ car will be used to reach a series of waypoints that are provided sequentially in the least amount of time. Overall, the method consists of deriving the set of differential equations and boundary conditions for the TOC problem using PMP. Given that the resulting differential equations for the TOC problem involve first-order differential equations it is proposed to use Galerkin's Weighted Residuals Method (GWRM) to solve the two-point boundary value problem and find the approximate solution of $\mathbf{X}^{\star}(t)$ and $\ \mathbf{u}^{\star}(t),$ for all $t\in[0, T^{\star}]$.

This manuscript's main contribution is the development of a framework for reliable solution of TOC problems using GWRM and Sequential Convex Programming (SCP) as opposed to the shooting method which does not always provide a solution to the equations that result from PMP, even when both methods are initialized using the same set of heuristics. Once the functions for the trajectory and controller have been found, the nominal trajectory can be tracked using MPC to provided a closed-loop control law and reject possible disturbances on the system. Furthermore, the TOC problem using GWRM is solved in a comparable time as the shooting method.


\section{RELATED WORK}

Although using FEM to solve optimal control problems is not a widespread approach due to its high computational cost, the current growth in computational power is leading to a reconsideration of FEM for this purpose.  One of the earliest uses of FEM for optimal control problems was developed by Neuman \& Sen \cite{Neuman1974WeightedControl} in 1974, where they implemented the weighted residual method (WRM) for a one dimensional system with linear dynamics using GWRM and the collocation WRM. The results showed that using only two basis functions to approximate the solution of the optimal control problem achieved a maximum of 0.9\% degradation of the cost compared with the optimal control solution.

Hodges \& Bless \cite{Hodges1991WeakProblems} in 1991 applied the FEM to two-state-dimensional first-order linear systems, solving both Fixed-Final-Time problems as well as Free-Final-Time problems. The results showed positive results with errors in the approximation of the solution that where a function of the square of the number of elements, and that required as few as four elements to produce accurate results. Becker \& Rannacher \cite{Becker2001AnMethods} implemented an adaptive element discretization using Galerkin's Finite Element Method to solve a second-order linear system for a minimum fuel consumption optimal control problem. The method proposed by Becker \& Rannacher \cite{Becker2001AnMethods} was later applied by Kraft \& Larsson \cite{Kraft2010TheEquations} to a problem of optimal control that involved the (linear) dynamics of a vehicle that comprised a higher dimensional state.

Singh \cite{Singh2010AControl} explored the use of FEM for optimal control problems with nonlinear dynamics. In his work, Sigh explored different families of basis functions to approximate the solutions of optimal control problems, highlighting which types of problems benefited from particular choices of basis functions. He also demonstrated the effectiveness of FEM for optimal control problems with discontinuous control.

More recent works that utilize different approaches to WRM include \cite{Porwal2021AConstraints}, \cite{Zhou2016FiniteEquation} and \cite{Fuhrer2023Least-squaresProblems}.


\section{APPROACH}

As mentioned previously, the proposed approach consist on three different components: finding the set of differential equations and boundary values of the TOC problem using PMP, solving the set of differential equations using GWRM and finally tracking the nominal trajectory using MPC. The problem statement for each of the components is provided in the following sections.

\subsection{Pontryagin’s Maximum Principle - PMP}

Given the state $\mathbf{X}: \Omega = [0, T] \rightarrow \mathbb{R}^n$, control input $\mathbf{u}: \Omega \rightarrow \mathcal{U} \subseteq \mathbb{R}^m$ and dynamics $f: \mathbb{R}^n \times \mathbb{R}^m \rightarrow \mathbb{R}^n$ we want to find an optimal trajectory ($\mathbf{X}^{\star}(t)$, $\mathbf{u}^{\star}(t)$, $T^{\star}$) such that the cost defined by the terminal cost $J_T: \mathbb{R}_{++} \times \mathbb{R}^n  \rightarrow \mathbb{R}$ and stage cost $J: \mathbb{R}^n \times \mathbb{R}^m \rightarrow \mathbb{R}$ is minimized (\ref{eq:TOC}). Since it is a TOC problem, $T$ is a free variable.

\begin{equation}
    \begin{aligned}
    \minimize_{\mathbf{X}(t), \mathbf{u}(t), T \geq 0} \quad & J_T(T, \mathbf{X}(T)) + \int_{\Omega} J(\mathbf{X}(t), \mathbf{u}(t)) \ dt \\
\subjectto \quad & \mathbf{\dot{X}}(t) = f(\mathbf{X}(t), \mathbf{u}(t)), \ \mathbf{u}(t) \in \mathcal{U}, \ \forall t \in \Omega \\
& \mathbf{X}(0) = \mathbf{X}_0, \quad \mathbf{X}(T) \in \mathcal{X}_T
    \end{aligned}
    \label{eq:TOC}
\end{equation}

Given the Hamiltonian $\mathcal{H}_{\eta}: \Omega \times \mathbb{R}^n \times \mathbb{R}^m \times \mathbb{R}^n \rightarrow \mathbb{R}$ (\ref{eq:H}), Pontryagin's Maximum Principle states that \cite{Clarke2013FunctionalControl}: Let ($\mathbf{X}^{\star}(t)$, $\mathbf{u}^{\star}(t)$, $T^*$) be a local minimizer for the problem with bounded control set $\mathcal{U}$, then there exist a scalar $\eta \in \{0, 1\}$ and an optimal adjoint state $\mathbf{\lambda}^{\star} :\Omega^{\star} = [0, T^{\star}] \rightarrow \mathbb{R}^n$ such that condition (\ref{eq:cond_1}) through (\ref{eq:cond_6}) are satisfied.

\begin{equation}
\mathcal{H}_\eta(t,\mathbf{X}, \mathbf{u}, \mathbf{\lambda}) := \mathbf{\lambda}^Tf(\mathbf{X}, \mathbf{u}) - \eta J(\mathbf{X}, \mathbf{u})
\label{eq:H}
\end{equation}

\begin{enumerate}
    \item \textit{Non-triviality:}
    \begin{equation}
    (\eta, \mathbf{\lambda}^{\star}(t)) \neq 0 
    \label{eq:cond_1}
    \end{equation}
    \item \textit{Adjoint equation:} 
    \begin{equation}
    -\dot{\mathbf{\lambda}}^{\star} = \nabla_x \mathcal{H}_\eta(t,\mathbf{X}^{\star}, \mathbf{u}^{\star}, \mathbf{\lambda}^{\star}), \ \forall t \in \Omega^{\star}
    \end{equation}
    \item \textit{Transversality:}
    \begin{equation}
    -\mathbf{\lambda}^{\star}(T^{\star}) - \eta \nabla J_T(T^{\star},\mathbf{X}^{\star}(T^{\star})) \perp_{\mathbf{X}^{\star}(T)} \in \mathcal{X}_T   
    \end{equation}
    \item \textit{Maximum Condition:}
    \begin{equation}
    \mathcal{H}_\eta(t,\mathbf{X}^{\star}, \mathbf{u}^{\star}, \mathbf{\lambda}^{\star}) = \sup_{\mathbf{u}\in\mathcal{U}}  
 \mathcal{H}_\eta(t,\mathbf{X}^{\star}, \mathbf{u}, \mathbf{\lambda}^{\star}), \ \forall t \in \Omega^*    
    \end{equation}
    \item \textit{Maximum Condition at the Boundary:}
    \begin{equation}
    \mathcal{H}_\eta(T^{\star},\mathbf{X}^{\star}, \mathbf{u}^{\star}, \mathbf{\lambda}^{\star}) = \left. \eta \frac{\partial J_T}{\partial T}\right|_{T^{\star},\mathbf{X}^{\star}(T^{\star})}
    \end{equation}
    \item \textit{(Weak) Maximum Condition:}
    \begin{equation}
    \nabla \mathcal{H}_\eta(t, \mathbf{X}^{\star}, \mathbf{u}^{\star}, \mathbf{\lambda}^{\star}) \perp_{\mathbf{u}^{\star}(t)} \mathcal{U}, \ \forall t \in \Omega^{\star}
    \label{eq:cond_6}
    \end{equation}
\end{enumerate}

\subsection{Galerkin's Weighted Residuals Method - GWRM}

The weighted residual method \cite{Galerkin1915RodsPlates}, \cite{Fletcher1984ComputationalMethods} is a generic class of methods developed to obtain approximate solutions to the differential equations of the form given by (\ref{eq:DiffEq}).\footnote{Because from PMP, $\mathbf{u}^{\star}(t)$ is a function of $\mathbf{X}^{\star}(t)$ and $\mathbf{\lambda}^{\star}(t)$, it is not explicitly shown in the operator $\mathcal{T}$.}.
\begin{equation}
    \mathcal{T}(\mathbf{X}^{\star}(t), \mathbf{\lambda}^{\star}(t)) - \mathbf{g}(t) = \mathbf{0} \quad \forall t \in \Omega^{\star}
    \label{eq:DiffEq}
\end{equation}

Where $\mathbf{X}^{\star}(t)$ and $\mathbf{\lambda}^{\star}(t)$ are the dependent and unknown functions and $\mathbf{g}: \Omega \rightarrow \mathbb{R}^{2n}$ is a known function. $\mathcal{T}$ denotes a differential operator (non-linear in general) involving the derivatives of $\mathbf{X}^{\star}(t)$ and $\mathbf{\lambda}^{\star}(t)$ which define the differential equations over the domain $\Omega^*$. Let $\hat{\mathbf{X}}^{\star}(t) \approx \mathbf{X}^{\star}(t)$ and $\hat{\mathbf{\lambda}}^{\star}(t) \approx \mathbf{\lambda}^{\star}(t)$ be an approximate solution of the differential equation. Therefore the residual $\mathbf{R}: \Omega^{\star} \rightarrow \mathbb{R}^{2n}$ is defined by (\ref{eq:Residual}). 

\begin{equation}
     \mathbf{R}(t) = \mathcal{T}(\hat{\mathbf{X}}^{\star}(t), \hat{\mathbf{\lambda}}^{\star}(t)) - \mathbf{g}(t) \quad \forall t \in \Omega^{\star}
     \label{eq:Residual}
\end{equation}

The FEM optimal control problem in an abstract way can be formulated as follows: let $\mathbf{\mathcal{W}} = \mathbf{\mathcal{W}}_1 \times \mathbf{\mathcal{W}}_2$ with $\mathbf{\mathcal{W}}_1 = \mathbf{\mathcal{W}}_2 = \{ \mathbf{w} = \left[w_1, \cdots, w_{n} \right]^T | \ w_i : \Omega \rightarrow \mathbb{R}, \ w_i \in \mathcal{C}^p ,\ \forall i = \{1, \cdots, n\} \}$ be normed vector spaces with $p \geq 0$; $\mathbf{\mathcal{S}} \subset \mathbf{\mathcal{W}}_1$ and $\mathbf{\mathcal{Q}} \subset \mathbf{\mathcal{W}}_2$ be affine spaces; and $\mathbf{\mathcal{V}}_1 \subset \mathbf{\mathcal{W}}_1$, $\mathbf{\mathcal{V}}_2 \subset \mathbf{\mathcal{W}}_2$ be vector spaces such that $\mathcal{V}_1$ and $\mathcal{V}_2$ are the direction of $\mathcal{S}$ and $\mathcal{Q}$ respectively. Defining $\mathbf{\mathcal{V}} = \mathbf{\mathcal{V}}_1 \times \mathbf{\mathcal{V}}_2$ and the smooth functional $\mathcal{F}( \ \cdot \ ; \ \cdot \ ): \mathcal{S} \times \mathcal{Q} \times \mathcal{V} \rightarrow \mathbb{R}^{2n}$ (\ref{eq:Funcitonal}) which is linear in its second argument\footnote{The symbol $\odot$ represents the Hadamard product.}, the FEM states: find $\mathbf{X}^{\star}(t) \in \mathcal{S} = \{ \mathbf{s} \in \mathcal{W}_1 \ | \ \mathbf{s}(0) = \mathbf{X}_0, \ \mathbf{s}(T^{\star}) \in  \mathcal{X}_T \}$ and $\lambda^{\star}(t) \in \mathcal{Q} = \{ \mathbf{q} \in \mathcal{W}_2 \ | \ \mathbf{q}(T^{\star}) = \lambda(T^{\star}) \}$ such that (\ref{eq:FEM}) is satisfied.

\begin{equation}
    \mathcal{F}( \mathcal{S}, \mathcal{Q} \ ; \mathcal{V} ) = \int_{\Omega^*} \mathbf{R}(t) \odot \mathbf{v}(t) \ dt
    \label{eq:Funcitonal}
\end{equation}
\begin{equation}
    \begin{split}
        \mathcal{F}(\mathbf{X}^{\star}(t), \lambda^{\star}(t) \ ; \ \mathbf{v}(t) ) = \mathbf{0} \\
        \forall \mathbf{v} \in \mathcal{V} = \{ \mathbf{v} \in \mathcal{V} \ | \ \mathbf{v}_1(0) = \mathbf{0}, \ \mathbf{v}(T^{\star}) = \mathbf{0} \}
    \end{split}
    \label{eq:FEM}
\end{equation}

Defining a set of basis functions $ \mathbf{w}_{1,h} = \mathcal{W}_{1,h} \subset \mathcal{W}_1$, $ \mathbf{w}_{2,h} = \mathcal{W}_{2,h} \subset \mathcal{W}_2$, $\mathcal{W}_{1,h} \times \mathcal{W}_{2,h} = \mathcal{W}_h \subset \mathcal{W}$ and using the same basis functions for both trial functions $\mathbf{s}_h \in \mathcal{S}_h \subset \mathcal{W}_{1,h}$, $\mathbf{q}_h \in \mathcal{Q}_h \subset \mathcal{W}_{2,h}$, and test functions $\mathbf{v}_h \in \mathcal{V}_h \subset \mathcal{W}_{h}$ we arrive to Galerkin's Weighted Residuals Method (GWRM) that states: find $\hat{\mathbf{X}}^{\star}(t) \in \mathcal{S}_h = \{ \mathbf{s}_h \in \mathcal{W}_{1,h} \ | \ \mathbf{s}_h(0) = \mathbf{X}_0, \ \mathbf{s}_h(T^{\star}) \in \mathcal{X}_T \}$ and $\hat{\lambda}^{\star}(t) \in \mathcal{Q}_h = \{ \mathbf{q}_h \in \mathcal{W}_{2,h} \ | \ \mathbf{q}_h(T^{\star}) = \hat{\lambda}(T^{\star}) \}$ such that (\ref{eq:GWRM}) is satisfied.

\begin{equation}
    \begin{split}
        \mathcal{F}(\hat{\mathbf{X}}^{\star}(t), \hat{\lambda}^{\star}(t) \ ; \ \mathbf{v}_h(t) ) = \mathbf{0} \\
        \forall \mathbf{v}_h \in \mathcal{V}_h = \{ \mathbf{v}_h \in \mathcal{V}_h \ | \ \mathbf{v}_{1,h}(0) = \mathbf{0}, \ \mathbf{v}_h(T^{\star}) = \mathbf{0} \}
    \end{split}
    \label{eq:GWRM}
\end{equation}

The approximate solution for $\mathbf{X}^{\star}(t)$ (\ref{eq:xhat}) and $\lambda^{\star}(t)$ (\ref{eq:lambdahat}) are linear combinations of the basis functions of the affine spaces $\mathcal{S}_h$ and $\mathcal{Q}_h$ respectively. $N$ is the number of basis functions that define the affine spaces $\mathcal{S}_h$ and $\mathcal{Q}_h$.

\begin{equation}
    \hat{\mathbf{X}}^{\star}(t) = \sum_{k=1}^{N} \mathbf{\alpha}_k \odot \mathbf{s}_{h,k}, \quad \mathbf{s}_h \in \mathcal{S}_h
    \label{eq:xhat}
\end{equation}

\begin{equation}
    \hat{\lambda}^{\star}(t) = \sum_{k=1}^{N} \mathbf{\beta}_k \odot \mathbf{q}_{h,k}, \quad \mathbf{q}_h \in \mathcal{Q}_h
    \label{eq:lambdahat}
\end{equation}

Because the set of differential equations are non-linear, the functional $\mathcal{F}$ is a non-linear function of the coefficients $\alpha_k \in \mathbb{R}^n$ and $\beta_k \in \mathbb{R}^n$ (\ref{eq:SetEqs}).

\begin{equation}
\mathbb{F}(\alpha_{1:N}, \beta_{1:N}, T^{\star}) = \begin{bmatrix} \mathcal{F}_1(\hat{\mathbf{X}}^{\star}(t), \hat{\lambda}^{\star}(t) \ ; \ \mathbf{v}_{h,1}(t) )\\
\vdots \\
\mathcal{F}_{N}(\hat{\mathbf{X}}^{\star}(t), \hat{\lambda}^{\star}(t) \ ; \ \mathbf{v}_{h,N}(t) )\\
\end{bmatrix} = 0
\label{eq:SetEqs}
\end{equation}

The objective is then two find the values of $\alpha_k \in \mathbb{R}^n$, $\beta_k \in \mathbb{R}^n$ and $T^{\star}$ that satisfy (16).

\begin{equation}
    \begin{aligned}
        \find_{\alpha_{1:N}, \beta_{1:N}, T^{\star}} & \\
\subjectto & \quad \mathbb{F}(\alpha_{1:N}, \beta_{1:N}, T^{\star}) = 0 \\
& \quad \alpha_1 = \mathbf{X}_0, \quad \alpha_{N} \in \mathcal{X}_T
    \end{aligned}
\label{eq:SCP_Nonlinear}
\end{equation}

Since \ref{eq:SCP_Nonlinear} is non-linear, an alternative to solve this problem is to linearize it to find the variation of the parameters using Sequential Convex Programming (SCP) over a trust region defined by $\rho_s \in \mathbb{R}_{++}$, $\rho_q  \in \mathbb{R}_{++}$ and $\rho_T  \in \mathbb{R}_{++}$, given initial values $\alpha_{1:N}^{(0)}, \ \beta_{1:N}^{(0)}, \ T^{\star(0)}$ (\ref{eq:SCP}).

\begin{equation}
    \begin{aligned}
        \minimize_{\delta \alpha_{1:N}, \delta \beta_{1:N}, \delta T^{\star}} & \quad ||\mathbf{b}^{(i)} + \mathbf{A}^{(i)} \left[\delta \alpha_{1:N} \ \ \delta \beta_{1:N} \ \ \delta T^{*} \right]^T || \\
\subjectto & \quad \delta \alpha_1 = \mathbf{0}, \quad \delta \beta_{N} = \mathbf{0} \\
& \quad \delta \alpha_{N} + \alpha^{(i)} \in \mathcal{X}_T, \quad ||\delta \alpha_{1:N}||_{\infty} \leq \rho_s \\
& \quad ||\delta \beta_{1:N}||_{\infty} \leq \rho_q, \quad ||\delta T^{\star}||_{\infty} \leq \rho_T
    \end{aligned}
\label{eq:SCP}
\end{equation}

Such that

\begin{equation}
    \mathbf{A}^{(i)} = \frac{\partial \mathbb{F}(\alpha_{1:N}^{(i)}, \beta_{1:N}^{(i)}, T^{\star(i)})}{\partial \alpha_{1:N}, \beta_{1:N}, T^{\star}}
\end{equation}
\begin{equation}
    \mathbf{b}^{(i)} = \mathbb{F}(\alpha_{1:N}^{(i)}, \beta_{1:N}^{(i)}, T^{\star(i)})
\end{equation}
\begin{equation}
    \alpha_{1:N}^{(i+1)} \leftarrow \alpha_{1:N}^{(i)} + \delta \alpha_{1:N}
\end{equation}
\begin{equation}
    \beta_{1:N}^{(i+1)} \leftarrow  \beta_{1:N}^{(i)} + \delta \beta_{1:N}
\end{equation}
\begin{equation}
    T^{\star(i+1)} \leftarrow  T^{\star(i)} + \delta T^{\star}
\end{equation}

The constraints on $\delta \alpha_1$, $\delta \alpha_{N}$ and $\delta \beta_{N}$ in (\ref{eq:SCP}) arise because the initial state $\mathbf{X}_0$, terminal set $\mathcal{X}_T$ and final state of the adjoint state $\lambda({T^{\star}})$ are known from the TOC problem and PMP.

\subsection{Model Predictive Control - MPC}

Finally, model predictive control (MPC) is applied to track the trajectory generated from the GWRM, where $\mathbf{P}_H \succeq 0$, $\mathbf{Q} \succeq 0$ and $\mathbf{R} \succ 0$ define the terminal and stage cost of the MPC (\ref{eq:MPC}).

\begin{equation}
    \begin{aligned}
    \minimize_{\mathbf{u}_{0:H-1}} \quad & ||(\mathbf{X}_H - \hat{\mathbf{X}}^{\star}_H)||_{\mathbf{P}_H} + \\
    & \sum_{k=0}^{H-1} \left[||\mathbf{X}_k - \hat{\mathbf{X}}^{\star}_k||_\mathbf{Q} + ||\mathbf{u}_k - \mathbf{\hat{u}}^{\star}_k||_\mathbf{R}\right] \\
    \subjectto \quad & \mathbf{X}_{k+1} = \mathbf{A}_{f,k}\mathbf{X}_k + \mathbf{B}_{f,k}\mathbf{u}(t) + \mathbf{c}_{f,k}\\
    & \mathbf{u}_k \in \mathcal{U}, \quad \forall k \in \{0, \cdots H-1 \} \\
      & \mathbf{X}_0 = \mathbf{X}(t), \quad \mathbf{X}_H \in \mathcal{X}_H
    \end{aligned}
    \label{eq:MPC}
\end{equation}

$\mathbf{A}_{f,k}$, $\mathbf{B}_{f,k}$ and $\mathbf{c}_{f,k}$ in (\ref{eq:MPC}) result from linearizing the dynamics around the nominal trajectory $\mathbf{\hat{X}^{\star}}_k, \  \mathbf{\hat{u}}^{\star}_k$. Notice that $||\mathbf{X} - \hat{\mathbf{X}}^{\star}||_\mathbf{Q} \triangleq (\mathbf{X} - \hat{\mathbf{X}}^{\star})^T\mathbf{Q}(\mathbf{X} - \hat{\mathbf{X}}^{\star})$.

\section{NUMERICAL ILLUSTRATION: DUBINS' CAR}

The proposed method will be tested on the Dubins' car with the system dynamics given by (\ref{eq:Dynamics}).

\begin{equation}
    \dot{x} = v \ cos(\theta), \quad \dot{y} = v \ sin(\theta), \quad \dot{\theta} = \omega
    \label{eq:Dynamics}
\end{equation}

Where $(x, y)$ is the position of the vehicle, $\theta$ is its heading angle, $v$ is its forward velocity, and $\omega$ is its angular velocity. Overall, the state and control input for this system are $\mathbf{X}:=\left[x, \ y, \ \theta\right]^T \in \mathbb{R}^3$ and $\mathbf{u} := \left[v, \ \omega\right]^T \in \mathbb{R}^2$. Let us assume that there are no constraints on the state space or the control set $\mathcal{U}$ and that the waypoints are provided sequentially, meaning that the trajectory needs to be generated online.  At any given time step we only know the starting state ($\mathbf{X}_0=\mathbf{0}$ without loss of generality) and the final state ($\mathbf{X}_T$), and the objective is to arrive to the final state in minimum time with as little control effort as possible. The cost function for the TOC problem is defined by (\ref{eq:CostFunc}), for some scalar values $\mu_{T} \in \mathbb{R}_{++}, \ \mu_{v}  \in \mathbb{R}_{++}, \ \mu_{\omega}  \in \mathbb{R}_{++}$.

\begin{equation}
    J_T = \mu_{T}T, \quad J = \int_{\Omega} \mu_{v} v^2 + \mu_{\omega} \omega^2 \ dt
    \label{eq:CostFunc}
\end{equation}

Applying PMP to this system yields the set of ordinary differential equations (\ref{eq:ODE1})-(\ref{eq:ODE5}), valid for all $\tau \in \bar{\Omega}$, and boundary values (\ref{eq:BC1})-(\ref{eq:BC3}). Because it is a free-time problem, $\Omega = [0, T]$ is variable and the problem is solved over a reference interval $\bar{\Omega} = [0, 1]$ parameterized by $\tau$. Notice that the variables $t$, $x$ and $y$ are cyclic and therefore the corresponding adjoint states $\lambda_t^{\star}$, $\lambda_x^{\star}$, $\lambda_y^{\star}$ are constant along the trajectory.

\begin{equation}
    \frac{dx^{\star}}{d\tau} = \frac{T^{\star}}{2\mu_{v}} \left(\lambda_x^{\star} \cos^2\theta^{\star} + \lambda_y^{\star} \cos\theta^{\star} \sin\theta^{\star} \right)
    \label{eq:ODE1}
\end{equation}
\begin{equation}
    \frac{dy^{\star}}{d\tau} = \frac{T^{\star}}{2\mu_{v}} \left(\lambda_x^{\star} \cos\theta^{\star} \sin\theta^{\star} + \lambda_y^{\star} \sin^2\theta^{\star} \right)
\end{equation}
\begin{equation}
    \frac{d\theta^{\star}}{d\tau} = \frac{T^{\star}}{2\mu_{\omega}} \lambda_{\theta}^{\star}
\end{equation}
\begin{equation}
\begin{aligned}
    \frac{d\lambda_{\theta}^{\star}}{d\tau} = \frac{T^{\star}}{2\mu_{v}} \left[(\lambda_x^{\star2} - \lambda_y^{\star2})\cos\theta^{\star}\sin\theta^{\star}\right.\\
    \left.+ \lambda_x^{\star} \lambda_y^{\star} (\sin^2\theta^{\star} - \cos^2\theta^{\star}) \right]
\end{aligned}
\end{equation}
\begin{equation}
    \frac{dt^{\star}}{d\tau} = T^{\star}
    \label{eq:ODE5}
\end{equation}
\begin{equation}
    \mathbf{X}_0 = \left[x^{\star}(0), \ y^{\star}(0), \ \theta^{\star}(0) \right]^T
    \label{eq:BC1}
\end{equation}
\begin{equation}
    \mathbf{X}_T = \left[x^{\star}(1), \ y^{\star}(1), \ \theta^{\star}(1) \right]^T
\end{equation}
\begin{equation}
    t^{\star}(0) = 0
    \label{eq:BC3}
\end{equation}

Furthermore, since $t$ is cyclic, the Hamiltonian is a constant of the trajectory (\ref{eq:Hamiltoniant}).

\begin{equation}
    \begin{split}
        \frac{\partial \mathcal{H}_\eta}{\partial t} = 0 \implies \frac{1}{4\mu_{v}}(\lambda_x^{\star}\cos\theta^{\star}+\lambda_y^{\star}\sin\theta^{\star})^2 + \\
        \frac{1}{4\mu_{\omega}}\lambda_\theta^{\star2} - \mu_{T} = 0, \quad \forall \tau \in \bar{\Omega}
    \end{split}
    \label{eq:Hamiltoniant}
\end{equation}

The optimal control law, $v^{\star}$, $\omega^{\star}$, are given by (\ref{eq:vstar}) and (\ref{eq:omegastar}) respectively.

\begin{equation}
    v^{\star} = \frac{1}{2\mu_{v}}(\lambda_x^{\star}\cos\theta^{\star} + \lambda_y^{\star}\sin\theta^{\star})
    \label{eq:vstar}
\end{equation}
\begin{equation}
\omega^{\star} = \frac{1}{2\mu_{\omega}}\lambda_\theta^{\star}
\label{eq:omegastar}
\end{equation}

To apply GWRM, let us define the partition of the domain into $n_{el}$ intervals and $N = n_{el}+1$ nodes as (\ref{eq:partition}).

\begin{gather}
    \bar{\Omega}^e = [\tau_e, \tau_{e+1}], \quad
    \bar{\Omega} = \bigcup_{e=1}^{n_{el}} \bar{\Omega}^e, \quad 
    \bigcap_{e=1}^{n_{el}} \bar{\Omega}^e = \{\tau_1, \cdots, \tau_{N} \} \nonumber\\
    0 = \tau_1 < \cdots  < \tau_{N} = 1
    \label{eq:partition}
\end{gather}

For simplicity, let all the sub intervals have the same length. Additionally, the vector spaces $\mathcal{W}_{1,h}$, (\ref{eq:W1}), and $\mathcal{W}_{2,h}$, (\ref{eq:W2}), are formed using piece-wise Lagrange polynomials $\varphi$ of degree one (see \nameref{Appendix A} for the definition of the Lagrange polynomials). Notice that because $\lambda_x$ and $\lambda_y$ are constants, the dimension of $\mathcal{W}_2$ is smaller than the dimension of $\mathcal{W}_1$. Also the differential equation associated with $t$ (\ref{eq:ODE5}), does not need to be solved with GWRM since it is implicitly incorporated in the problem.

\begin{equation}
    \begin{split}
        \mathcal{W}_{1,h} = \text{span}\{ \varphi_1, \cdots, \varphi_{N}\} \times   \text{span} \{\varphi_1, \cdots, \varphi_{N}\} \\ \times \text{span}\{ \varphi_1, \cdots, \varphi_{N}\}\\
    \end{split}
    \label{eq:W1}
\end{equation}
\begin{equation}
    \mathcal{W}_{2,h} =  \text{span}\{ \varphi_1, \cdots, \varphi_{N}\}
    \label{eq:W2}
\end{equation}

Therefore the affine vector spaces $\mathcal{S}_h$ and $\mathcal{Q}_h$ and test functions $\mathbf{v}_{h,i}$ for all $i=\{1, \cdots, N\}$  are defined by (\ref{eq:AffineSub}).

\begin{equation}
    \begin{aligned}
        \mathcal{S}_h = \{ \mathbf{s}_h \in \mathcal{W}_{1,h} \ | \ \mathbf{s}_h(0) = \mathbf{X}_0, \ \mathbf{s}_h(1) = \mathbf{X}_T \}\\
    \mathcal{Q}_h = \{ \mathbf{q}_h \in \mathcal{W}_{2,h} \}\\
    \mathbf{v}_{h,i} = [\varphi_i, \ \varphi_i, \ \varphi_i, \ \varphi_i]^T, \quad \forall i=\{1, \cdots, N\}
    \end{aligned}
    \label{eq:AffineSub}
\end{equation}

With these definitions, the SCP in (\ref{eq:SCP2}) can be solved with the use of CVXPY \cite{Agrawal2018AProblems}, \cite{Diamond2016CVXPY:Optimization} to find the coefficients $\beta_{1:N} \equiv \{ \lambda_x^{\star}, \lambda_y^{\star}, \beta_{1:N}^{\lambda_{\theta}^{\star}} \}$, $\alpha_{1:N} = \{\alpha_{1:N}^{x^{\star}}, \ \alpha_{1:N}^{y^{\star}},\ \alpha_{1:N}^{\theta^{\star}}\}$ and $T^{\star}$ that form the GWRM solution. Because the Hamiltonian is known to be a constant of the trajectory, it can be incorporate in the SCP to better inform the search of the solution.

\begin{equation}
    \begin{aligned}
        \minimize_{\delta \alpha_{1:N}, \delta \beta_{1:N}, \delta T^{\star}} & \quad ||\mathbf{b}^{(i)} + \mathbf{A}^{(i)} \left[\delta \alpha_{1:N} \ \ \delta \beta_{1:N} \ \ \delta T^{*} \right]^T ||_1 \\
        & \quad + ||\mathbf{d}^{(i)} + \mathbf{C}^{(i)} \left[\delta \alpha_{1:N} \ \ \delta \beta_{1:N} \right]^T ||_1 \\
\subjectto & \quad \delta \alpha_1 = \delta \alpha_{N} = \mathbf{0}, \quad ||\delta \alpha_{1:N}||_{\infty} \leq \rho_s \\
& \quad ||\delta \beta_{1:N}||_{\infty} \leq \rho_q, \quad ||\delta T^{\star}||_{\infty} \leq \rho_T
    \end{aligned}
    \label{eq:SCP2}
\end{equation}
\begin{equation}
    \mathbf{C}^{(i)} = \begin{bmatrix}\frac{\partial \mathbb{\mathcal{H}_\eta}(\tau_1, \alpha_{1:N}^{(i)}, \beta_{1:N}^{(i)})}{\partial \alpha_{1:N}} & \frac{\partial \mathcal{H}_\eta(\tau_1, \alpha_{1:N}^{(i)}, \beta_{1:N}^{(i)})}{\partial \beta_{1:N}} \\ \vdots & \vdots \\ \frac{\partial \mathbb{\mathcal{H}_\eta}(\tau_N, \alpha_{1:N}^{(i)}, \beta_{1:N}^{(i)})}{\partial \alpha_{1:N}} & \frac{\partial \mathcal{H}_\eta(\tau_N, \alpha_{1:N}^{(i)}, \beta_{1:N}^{(i)})}{\partial \beta_{1:N}}
    \end{bmatrix}
\end{equation}

\begin{equation}
    \mathbf{d}^{(i)} = \begin{bmatrix}
    \mathbb{\mathcal{H}_\eta}(\tau_1, \alpha_{1:N}^{(i)}, \beta_{1:N}^{(i)}) \\ \vdots \\
    \mathbb{\mathcal{H}_\eta}(\tau_N, \alpha_{1:N}^{(i)}, \beta_{1:N}^{(i)})
    \end{bmatrix}
\end{equation}

The linearization required in the SCP (\ref{eq:SCP2}) is performed with the use of JAX \cite{Bradbury2018JAX:Programs}. The SCP is ran for a maximum number of iterations or until the variation of the parameters are below a tolerance $\varepsilon$, i.e. $||\begin{bmatrix}\delta \alpha_{1:N} & \delta \beta_{1:N} & \delta T^{*} \end{bmatrix}^T||_{\infty} \leq \varepsilon$. To initialize the SCP, as well as the shooting method for comparison, the following heuristics are used:
\begin{itemize}
    \item The velocity is a function of $\lambda_x^{\star}$ and $\lambda_y^{\star}$, so by intuiting the sign of the velocity $v^{\star(0)}$ (i.e. the system starts moving forward or backward) and given $\theta^{\star}(0)$, then the sign of $\lambda_x^{\star(0)}$ and $\lambda_y^{\star(0)}$  can be inferred from (\ref{eq:vstar}).  The magnitude could be set to some arbitrary value, for example 1.
    \item The angular velocity (\ref{eq:omegastar}) is a function of $\lambda_{\theta}^{\star}$, so by intuiting the sign of the angular velocity $\omega^{\star(0)}$ (i.e. the system starts turning left or right), then $\text{sign}(\lambda_{\theta}^{\star(0)}) = \text{sign}(\omega^{\star(0)})$ and the magnitude could be set to some arbitrary value, for example 1.
    \item The optimal time $T^{\star(0)}$ can be initialized as (\ref{eq:InitT2}) by assuming the system moves with a constant velocity from state $\mathbf{X}_0$ to $\mathbf{X}_T$, ignoring turning (\ref{eq:InitT1}).
\end{itemize}

\begin{equation}
    \begin{split}
        \mu_T T + \int_0^T \mu_v v^2 \approx \\ \mu_T T + \mu_v\frac{||[x(T)-x(0), \ y(T)-y(0)]^T||_2^2}{T}
    \end{split}
    \label{eq:InitT1}
\end{equation}

\begin{equation}
    T^{\star(0)} \leftarrow \sqrt{\frac{\mu_v}{\mu_T}||[x(T)-x(0), \ y(T)-y(0)]^T||_2^2}
     \label{eq:InitT2}
\end{equation}

The GWRM, different from the shooting method, needs not only a initialization of the adjoint state at the beginning of the trajectory $\mathbf{\lambda}^{\star(0)}(0)$ and the optimal time $T^{\star(0)}$, but also a trajectory (not necessarily dynamically feasible) of the state and adjoint state, $\mathbf{X}^{\star(0)}(\tau_{1:N}) \equiv \alpha_{1:N}^{(0)}$, $\lambda^{\star(0)}(\tau_{1:N}) \equiv \beta_{1:N}^{(0)}$. This is achieved using Bézier Curves (see \nameref{Appendix B}).

\begin{figure*}
    \centering
    \includegraphics[width=0.68\textwidth]{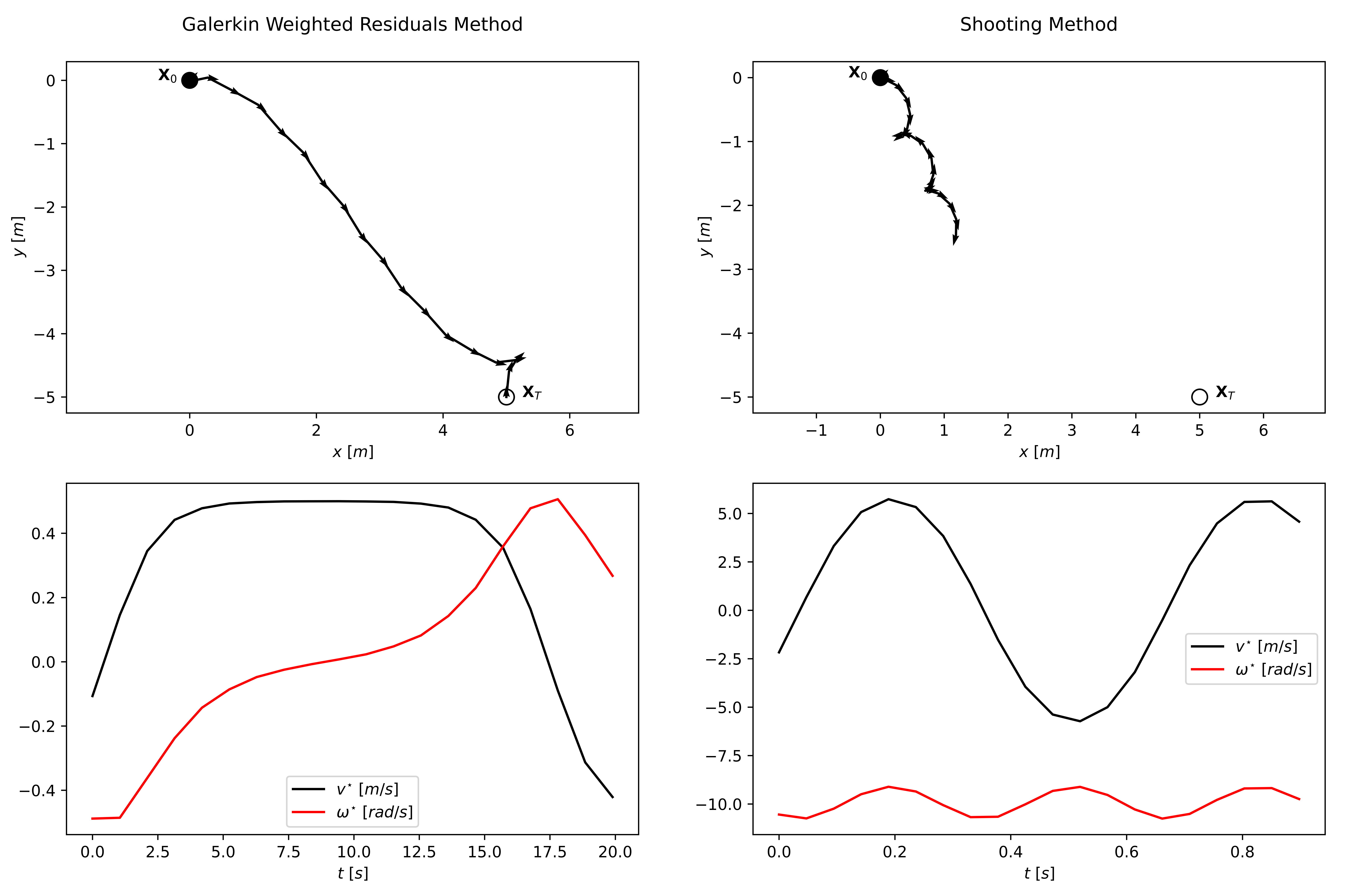}
    \caption{Comparison of time-optimal trajectory using GWRM (left) and the shooting method (right) for $\mathbf{X}_0 = [0, 0, \frac{\pi}{4}]$ and $\mathbf{X}_T = [5, -5, \frac{\pi}{2}]$}
    \label{fig:Example 2}
\end{figure*}

Fig. \ref{fig:Example 1} shows the performance of GWRM against a trajectory that converges using the shooting method. The GWRM is able to find the solution within 7 seconds, whereas the shooting methods finds the solution within 2 seconds. Fig. \ref{fig:Example 2} shows a trajectory with different initial and final state where the GWRM is able to find a trajectory within 25 seconds, whereas the shooting method is unable to converge. (See the parameters used for Fig. \ref{fig:Example 1} and Fig. \ref{fig:Example 2} in \nameref{Appendix C}).

\begin{figure*}
\centering
\includegraphics[width=0.665\textwidth]{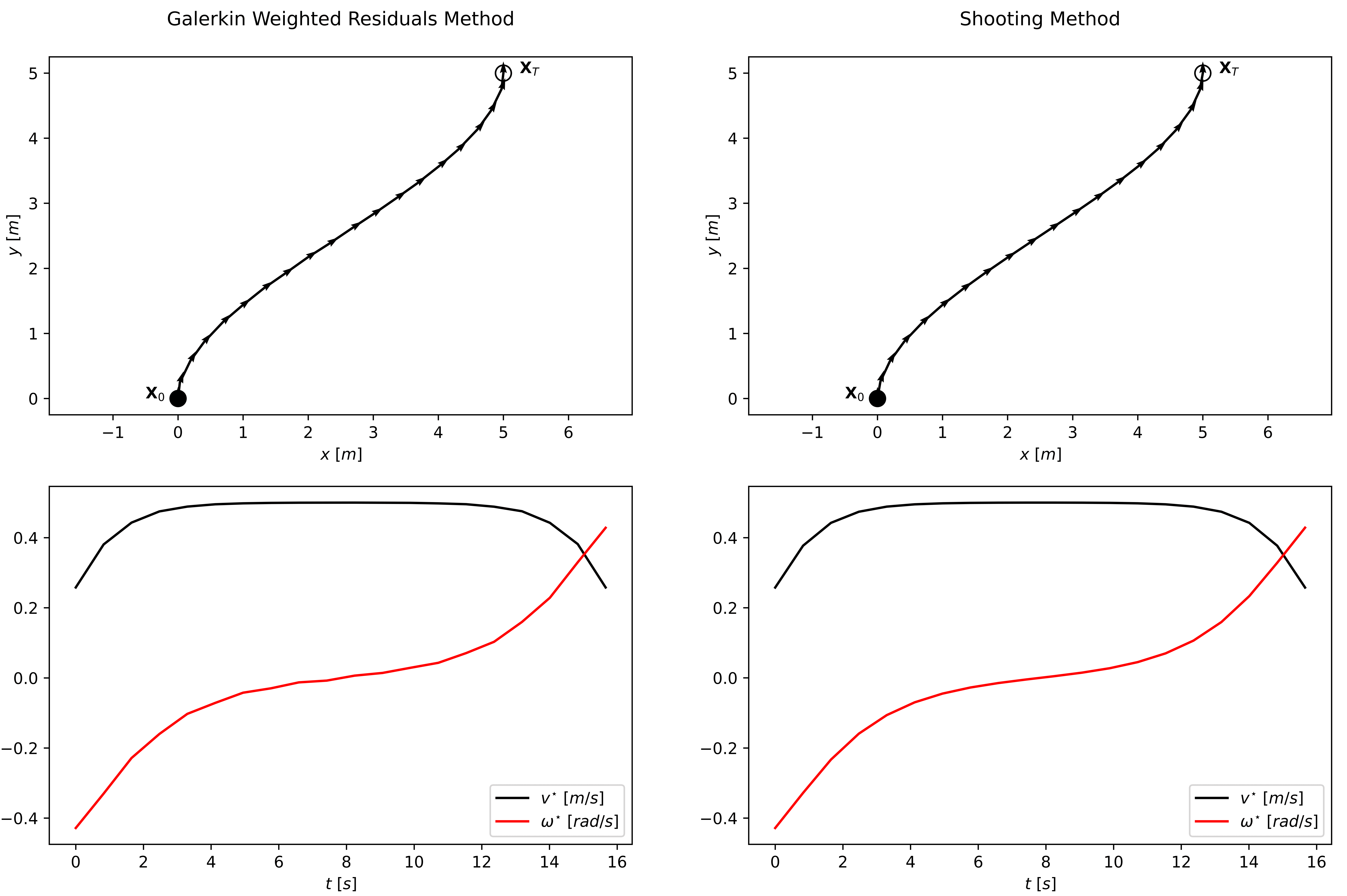}
\caption{Comparison of time-optimal trajectory using GWRM (left) and the shooting method (right) for $\mathbf{X}_0 = [0, 0, \frac{\pi}{2}]$ and $\mathbf{X}_T = [5, 5, \frac{\pi}{2}]$}
\label{fig:Example 1}
\end{figure*}

\begin{figure*}
    \centering
    \includegraphics[width=0.67\textwidth]{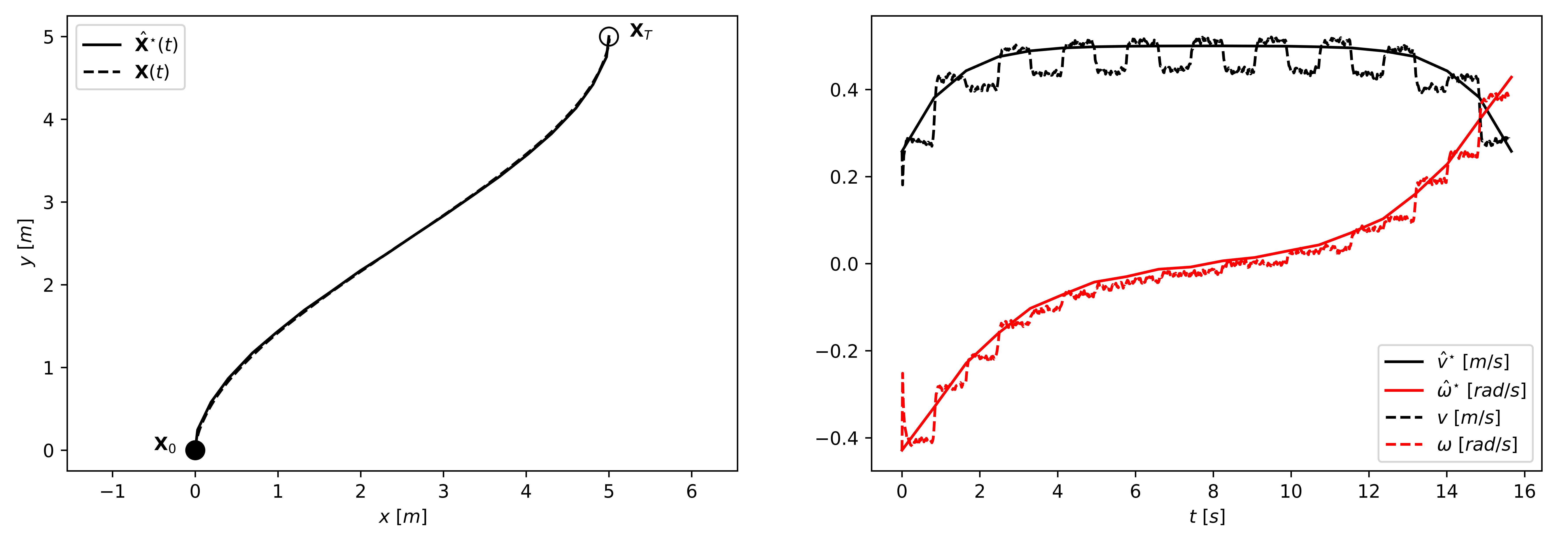}
    \caption{Resulting MPC trajectory $\mathbf{X}(t)$ and control actions $v(t)$, $\omega(t)$ after tracking TOC nominal trajectory $(\hat{\mathbf{X}}^{\star}(t), \hat{v}^{\star}(t), \hat{\omega}^{\star}(t), \hat{T}^{\star})$ obtained using GWRM.}
    \label{fig:MPC}
\end{figure*}

Finally, once the nominal trajectory $(\hat{\mathbf{X}}^{\star}(t), \ \hat{v}^{\star}(t), \ \hat{\omega}^{\star}(t), \ \hat{T}^{\star})$ is found using GWRM, MPC is used to track the trajectory, rejecting possible disturbances and providing a closed-loop control around the open-loop control generated from the TOC problem. Fig. \ref{fig:MPC} shows the MPC tracking with control frequency $\mathit{f} = 1000/T^{\star}$ Hz where the dynamics have added IID noise with covariance matrix $\Sigma = \delta t^{2}\times10^{-3}\mathbf{I}$ where $\delta t$ is the time step of the MPC (See the parameters used for Fig. \ref{fig:MPC} in \nameref{Appendix C}).

\section{CONCLUSIONS}

This paper demonstrates the use of Finite Element Methods to solve time optimal control problems. Given the dynamics of the system and a cost function, the system of ordinary differential equations that enables one to find the time-optimal control trajectory can be found using Pontryagin's Maximum Principle (PMP). These set of differential equations, that in general are non-linear, are a two-boundary value problem that are typically solved using shooting methods, which heavily depend on the initialization of the method and may not always converge. Overcoming this dependency on initialization can be achieved by incorporating both boundaries into the method. Galerkin's Weighted Residuals Method (GWRM) and Sequential Convex Programming (SCP) are employed for this purpose. The method is validated using as motivation the kinematics of a Dubins' Car, showing that when both shooting method and GWRM converge, both produce a very similar trajectories in comparable time. Furthermore, it is demonstrated that in some cases, despite similar heuristic initialization of both methods, the shooting method may fail to produce a time-optimal trajectory, while GWRM is capable of doing so.

After generating the time-optimal trajectory, Model Predictive Control (MPC) can be employed to track the open-loop optimal trajectory in a closed-loop fashion, effectively mitigating potential disturbances in the system dynamics.

Potential future research directions may involve: using Lagrange polynomials of higher degree to provide smoother trajectories, including adaptive discretizations of the reference interval for a more accurate representation of the state trajectory and control functions, and optimizing the implementation of the method to reduce the computational time required for trajectory computation, particularly exploring the use of quadrature rules.

\section*{APPENDIX A}\label{Appendix A}

The piece-wise Lagrange polynomials $\varphi$ of degree one for the partition of the domain used for the Dubins' car problem are defined by (\ref{eq:InteriorOmega}) in the interior of $\Bar{\Omega}$ (i.e. for all $e = \{2, \cdots, n_{el}\}$), and  (\ref{eq:LeftOmega}), (\ref{eq:RightOmega}) in the boundaries. Notice that $\varphi \in \mathcal{C}^0$ over the domain $\Bar{\Omega}$.

\begin{equation}
    \varphi_e = \begin{cases}
    \frac{\tau - \tau_{e-1}}{\tau_e-\tau_{e-1}} & ,\tau \in \Bar{\Omega}^{e-1} \\
    \frac{\tau - \tau_{e+1}}{\tau_e-\tau_{e+1}} & ,\tau \in \Bar{\Omega}^e \\   0 & ,\tau \in \bar{\Omega} \backslash (\Bar{\Omega}^{e-1} \cup \Bar{\Omega}^e )
    \end{cases}
    \label{eq:InteriorOmega}
\end{equation}

\begin{equation}
    \varphi_1 = \begin{cases}
    \frac{\tau - \tau_2}{\tau_1-\tau_2} & ,\tau \in \Bar{\Omega}^1 \\   0 & ,\tau \in \bar{\Omega} \backslash \Bar{\Omega}^1
    \end{cases}
    \label{eq:LeftOmega}
\end{equation}

\begin{equation}
    \varphi_{N} = \begin{cases}
    \frac{\tau - \tau_{N-1}}{\tau_{N}-\tau_{N-1}} & ,\tau \in \Bar{\Omega}^{n_{el}} \\   0 & ,\tau \in \bar{\Omega} \backslash \Bar{\Omega}^{n_{el}}
    \end{cases}
    \label{eq:RightOmega}
\end{equation}

\section*{APPENDIX B}\label{Appendix B}

A Bézier curve is a parametric curve, parameterized by $\tau$ over the interval $[0,1]$ and a set of discrete control points $\mathbf{P}_i$. Bézier curves $\mathbf{B}(\tau)$ are defined as a linear combination of Bernstein basis polynomials $b_{i,n}$ of degree $n$.

\begin{equation}
    \mathbf{B}(\tau) = \sum_{i=0}^n b_{i,n}(\tau)\mathbf{P}_i = \sum_{i=0}^n \binom{n}{i}\tau^{i}(1-\tau)^{n-i}\mathbf{P}_i
\end{equation}

To initialize the trajectory of the Dubin's car, four control points are needed: the two end-points and two control points that enforce the heading angles at the beginning and end of the trajectory. Therefore the initial trajectory for $x^{\star(0)}(\tau)$ and $y^{\star(0)}(\tau)$ is generated using cubic Bézier curves, and the trajectory for $\theta^{\star(0)}(\tau)$ is implicitly defined by these two.

\section*{APPENDIX C}\label{Appendix C}

Parameters used to generate the plots in Fig. \ref{fig:Example 1} and Fig. \ref{fig:Example 2}: 
$n_{el} =19$, $N=20$, $\eta=1$, $\mu_T=0.25$, $\mu_{v}=1$, $\mu_{\omega}=1$, $\rho_s = \rho_q = \rho_T = 1$, $\varepsilon=1\times10^{-2}$, $\vartheta=1\times10^{-4}$.

Parameters used to generate the plots in Fig. \ref{fig:MPC}: $\delta t = \mathit{f}^{-1} = 0.0156 \ s$, $H=5$, $\mathbf{P}_H = \mathbf{I}$, $\mathbf{Q} = \mathbf{I}$, $\mathbf{R} = 1\times10^{-2}\mathbf{I}$.


\section*{ACKNOWLEDGMENT}

The author would like to thank Adrian J. Lew for his insights regarding the nonlinear finite element method.



\end{document}